\documentclass[aps,pra,twocolumn,groupaddress,showpacs,longbibliography]{revtex4-1}
\usepackage{amsmath,amssymb,graphicx,stmaryrd}
\usepackage{mathtools}
\usepackage[utf8]{inputenc}
\usepackage[T1]{fontenc}
\usepackage{abraces}
\usepackage{tabularx}

\IfFileExists{newtxtext.sty}
   {\usepackage{newtxtext,newtxmath}}
   {\IfFileExists{stix.sty}
      {\usepackage{stix}}
      {\IfFileExists{mathptmx.sty}
      {\usepackage{mathptmx}}{} } }

\usepackage{textcomp}

\usepackage{bm}
\usepackage{bbold}
\usepackage{makecell}
\usepackage{float}

\newcommand{\ud}{\hspace{0.1em}(\hspace{0.2em})\hspace{0.1em}}

\renewcommand{\vec}[1]{\bm{#1}}
\DeclareMathOperator{\sgn}{sgn}
\newcommand{\bra}[1]{\langle#1\rvert}
\newcommand{\ket}[1]{\lvert#1\rangle}

\newcommand{\expectation}[1]{\langle#1\rangle}

\usepackage[dvipsnames]{xcolor} % for color definition
\definecolor{blue}{rgb}{0.0, 0.38, 0.68}% '#0060ad'
\definecolor{red}{rgb}{0.87, 0.09, 0.12} %#dd181f
\definecolor{green}{rgb}{0.0, 0.50, 0.0} %#008000
\definecolor{brown}{rgb}{0.62, 0.50, 0.44} %#9f8170

\newcommand{\sztot}{S^z_\text{tot}}
\newcommand{\totsz}[1]{S^z_\text{tot}(\text{#1})}

\newcommand{\up}{\hspace{0.1em}(\hspace{0.1em}}
\newcommand{\dn}{\hspace{0.1em})\hspace{0.1em}}
\newcommand{\upquiet}{\hspace{0.1em}{\color{gray}(}\hspace{0.1em}}
\newcommand{\dnquiet}{\hspace{0.1em}{\color{gray})}\hspace{0.1em}}
\newcommand{\eo}{\hspace{0.1em}{\color{green}\lceil}\hspace{0.1em}}     % excited bond open
\newcommand{\ec}{\hspace{0.1em}{\color{green}\rceil}\hspace{0.1em}} % excited bond close
\newcommand{\deo}{\hspace{0.1em}{\color{blue}\lfloor}\hspace{0.1em}}     % excited bond open
\newcommand{\dec}{\hspace{0.1em}{\color{blue}\rfloor}\hspace{0.1em}} % excited bond close
\newcommand{\mo}{\hspace{0.1em}{\color{red}\langle}\hspace{0.1em}} % mismatch bond open
\newcommand{\mc}{\hspace{0.1em}{\color{red}\rangle}\hspace{0.1em}} % mismatch bond close

\usepackage{natbib}
\usepackage{setspace}

\pdfoutput=1
\usepackage{color}
\definecolor{LinkColor}{rgb}{0.256,0.439,0.588}
\usepackage{hyperref}
\hypersetup{
   pdfauthor={Khagendra Adhikari and K. S. D. Beach},
   pdftitle={Tunable quantum spin chain with three-body interactions},
   pdfsubject={ED and DMRG study of the Fredkin spin chain},
   pdfkeywords={Fredkin,} {quantum magnetism,} {phase transitions,} {competing ordered phases,} {ferromagnetism,} {antiferromagnetism,} {dimerization},
   colorlinks=true,
   citecolor=LinkColor,
   linkcolor=LinkColor,
   urlcolor=LinkColor
   }

\begin{document}

\title{Tunable quantum spin chain with three-body interactions}

\author{Khagendra Adhikari}
\email[Electronic address:\ ]{kadhikar@go.olemiss.edu}
\affiliation{Department of Physics and Astronomy, The University of Mississippi, University, Mississippi 38677, USA}

\author{K. S. D. Beach}
\email[Electronic address:\ ]{kbeach@olemiss.edu}
\affiliation{Department of Physics and Astronomy, The University of Mississippi, University, Mississippi 38677, USA}

\date{November 15, 2020}

\begin{abstract}
We introduce a generalization of the Fredkin spin chain with tunable three-body interactions expressed in terms of conventional spin-half operators. Of the model's two free parameters, one controls the preference for Ising antiferromagnetism  and the other controls the strength of quantum fluctuations. In this formulation, the so-called $t$-deformed model (an exactly solvable, frustration-free Hamiltonian) lives on a unit circle centered on the origin of the phase diagram. The circle demarcates the boundary between ferromagnetic order inside and various antiferromagnetic phases outside. Throughout most of the non-ferromagnetic parts of the phase diagram, the ground state has Dyck word form:
i.e., all contributing spin configurations exhibit perfect matching and nesting of spin up and spin down. The exceptions are two regions in which Dyck word mismatches are energetically favorable. We remark that in those regions the energy level spacing can be exponentially small in the system size. It is also the case that exact diagonalization reveals 
a highly idiosyncratic energy spectrum, presumably because the
hard spin twist at the chain ends induces strong incommensurability effects 
on the bulk system when the chain length is small.
As a convergence check, we benchmark our DMRG results to near-double-precision floating-point accuracy against analytical results at exactly solvable points and against exact diagonalization results for small system sizes across the entire parameter space.
\end{abstract}
%@@@@@@@@@@@@@@@@@@@@@@@@@
\maketitle
 %@@@@@@@@@@@@@@@@@@@@@@@@@
\section{Introduction} \label{Intro}
Quantum spin chains were initially developed as toy models in the early days of quantum mechanics, but by the 1960s they had been realized experimentally in transition metal salts~\cite{Haseda-Physica-61,Flippen-JCP-63,Wagner-PL-64}. Many other fascinating material examples were subsequently discovered~\cite{Hone-PRB-74,Borsa-PRB-74,Motoyama-PRL-96,Bitko-PRL-96,Coldea-Science-10,Breunig-SA-17}. In spite of their simplicity, quantum spin chains exhibit complex properties and behaviors, such as magnetism~\cite{Zaliznyak-PRL-99,auerbach2012interacting}, scale-free criticality~\cite{Kojima-PRL-97}, quantum phase transitions~\cite{Bursill-JPCM-95,Jafari-PRB-07,Wierschem-MPLB-14,sachdev-2011}, topological order~\cite{Hirano-PRB-08,Pollmann-PRB-12}, short- and long-range correlations~\cite{Lieb-Annals-61}, and entanglement~\cite{Latorre-QIC-04,Keating-PRL-05}. This subject has reached a broader audience following the 2016 Nobel prize, which was awarded in part for Haldane's work to elucidate the topological origin of the divergent behavior of integer- and half-integer-spin chains~\cite{PhysRevLett.50.1153,RevModPhys.89.040502}. Research
in this area continues to be spurred by the search for new theoretical insights~\cite{Lieb-Annals-61,Affleck-PRB-87,Sirker-PRB-11,Patil-PRB-17} and by
the possibility of technological applications in spintronics~\cite{Hirobe-NP-17}, quantum communication~\cite{Bose-PRL-03,Barjaktarevic-PRL-05}, quantum computing~\cite{Zhou-PRL-02,Bartlett-PRL-10}, quantum simulations~\cite{Simon-Nature-11}, and quantum sensors~\cite{RevModPhys.91.041001}.

Frustration-free quantum spin chains with local three-body interactions are relatively recent discoveries, but they have generated great excitement and have already been studied extensively~\cite{Movassagh-PNAS-16,Ramis-Fredkin-Gap,Bravyi-PRL-12,Movassagh-JMP-17,DellAnna-PRB-16,Chen-PRB-17,Chen-JPA-17,Zhang-PNAS-17,Zhang-JMP-17,Levine-JPA-17,Udagawa-JPA-17,Padmanabhan-QIP-18,Deformed-spin-chain,PhysRevB-Adhikari-Beach,PhysRevB-tensor-network,SciPost-Long-distance-entanglement,Reneyi-Entropy-Korepin,pair-flip-arXiv-18,Salberger-RevMathPhys-16,PhysRevB.96.180404}. 
As is true for the original Motzkin model, 
the ground state of the Fredkin spin chain is known exactly. Despite being described by a local short-range Hamiltonian, the ground state exhibits robust nonlocal behavior, including long-distance entanglement~\cite{PhysRevB.96.180404} and violation of the cluster decomposition property~\cite{DellAnna-PRB-16}. Furthermore, the entanglement entropy grows as the square root of system size, putting to rest the folk wisdom that a Hamiltonian with local interactions must either obey the area law for a gapped system or deviate by at most logarithmic corrections for the gapless system.

The Fredkin spin chain~\cite{Salberger-RevMathPhys-16} is a spin-half chain segment subject to three-body correlated-exchange interactions and twisted boundary conditions. Its three-body interactions are structured such that a spin-singlet projector between adjacent spins is operative or not based on the spin state of a neighboring third site. The model is frustration free, and its ground state (GS) wave function is known to be an equal-weight superposition of all spin configurations of Dyck word form. This is possible because the interactions are in delicate balance. Various models~\cite{PhysRevB-Adhikari-Beach, Chen-JPA-17, Deformed-spin-chain, Zhang-JMP-17,Ramirez-JSM-15} have been proposed that continuously deform the Fredkin model away from this specially tuned point. In Refs.~\onlinecite{Chen-JPA-17} and \onlinecite{PhysRevB-Adhikari-Beach}, the model is changed to allow for a single tuning parameter that controls the strength and nature of the coupling to the third site. In the two extreme limits, the model reduces to the conventional Heisenberg models with ferromagnetic and antiferromagnetic two-body interactions. A different approach is to modify the two-site projector away from its spin-singlet form, as in Refs.~\onlinecite{Zhang-JMP-17} and \onlinecite{Deformed-spin-chain}. See the brief review in Appendix~\ref{Append:t-deformed-model}. 

The authors of Ref.~\onlinecite{Zhang-JMP-17} show that a Fredkin-like model can remain frustration-free while still allowing for an independent tuning parameter at each site that modifies the nature of the local projector (admixing singlet and triplet components). In the uniform case, referred to as the $t$-deformed Fredkin spin chain~\cite{Deformed-spin-chain}, an up-down pair of adjacent spins $\ket{\up\dn} = \ket{\hspace{0.1em}\uparrow\hspace{0.1em}\downarrow\hspace{0.1em}}$ moves passed its nested third neighbor as per $\upquiet\up\dn \Leftrightarrow \up\dn\upquiet$ or $ \up\dn\dnquiet \Leftrightarrow \dnquiet\up\dn$. These reconfigurations occur with different probability amplitudes that are functions of the real-valued $t$. The leftward and rightward motion of short matching pairs is symmetric at the Fredkin point ($t=1$). The quantum fluctuations freeze out entirely when $t$ approaches $0$ or $\pm\infty$; in these extreme limits, the Hamiltonian is semiclassical and the ground state is a pure product state.

 The ground state of the $t$-deformed model is the area-weighted superposition of all Dyck paths. Each weight goes as $t^A$, where $A = \tfrac{1}{2}\sum_{j=1}^Nh_j$ is the area under the spin configuration's height profile, $h_i = \sum_{j=1}^i \sigma_j^z$. The single maximum-area configuration $\up\up\up\up \cdots \dn\dn\dn\dn$ dominates as $\lvert t\rvert \to \infty$; the minimum-area configuration $\up\dn\up\dn \cdots \up\dn\up\dn$ dominates as $t\to 0$.
For $0 < \lvert t \rvert < 1$, the ground state favors Ising antiferromagnetic order (z AFM) and the excitations are gapped.
For $1 < \lvert t \rvert < \infty$, the ground state is featureless, and the excitation gap closes exponentially in the system size. Correlations in the easy
(xy) plane are either ferromagnetic ($\sgn t = t/\lvert t \rvert = +1$) or antiferromagnetic ($\sgn t = -1$).
\begin{figure}
\centering
	\includegraphics[width=0.85\linewidth]{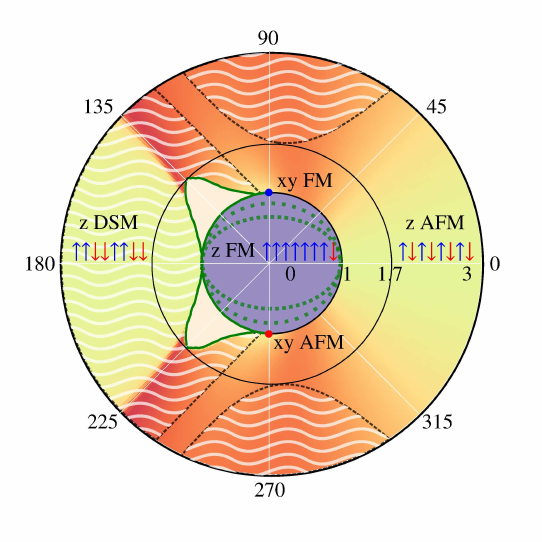}
	\caption{\label{FIG:phase-diagram-t-deformed} The diagram shows the various phases of the proposed deformed Fredkin model, plotted in the $\eta$--$\gamma$ plane. The false color background (applied outside the unit circle and outside the two domes) is based on DMRG measurements of dimer order for the $N = 60$ system; the dark red color implies stronger order, following the same scale used in Fig.~\ref{FIG:phases-60}(d). The numerical labels correspond to polar coordinates. Labels at the rim of the outermost circle mark the angle in degrees, and interior labels mark the radii. The patterns of up and down arrows describe the spin states of the classical ground state in the vicinity of the horizontal axis. The text labels denote regions characterized by Ising ferromagnetic order (z FM), Ising antiferromagnetic order (z AFM), doubly staggered Ising magnetic order (z DSM), xy-directed ferromagnetic correlations (xy FM), and xy-directed antiferromagnetic correlations (xy AFM). The relevant order parameters are defined in Sec.~\ref{SEC:t-deformed-methods}. The ground state is z FM inside the unit circle, favoring the fully polarized (except the rightmost down spin) state inside the smallest ellipse and stepping down across each dotted green line. Unlike all diagonal measurements, the off-diagonal measurements are antisymmetric about the horizontal diagonal axis. For example, the filled blue circle represents xy-directed FM at the top and the filled red circle denotes xy-directed AFM at the bottom on the unit circle. A distinct hyperbolic region on the left favors fully polarized z DSM. In much of the phase diagram, z~AFM and dimer order coexist, with dimer order dominating when $\gamma \gg \eta$ and vice versa. The regions with a unique non-Dyck-form ground state  appear as two cat's ears domes, bounded by a solid green line. The regions marked with wavy white hatching and enclosed by dotted black lines indicate where the first excited state is $\totsz{\text{ES}} = 0$ in character. The system is gapped everywhere except along the green (solid and dotted) lines.} 
\end{figure}

We consider a further generalization in terms of two parameters $\eta$ and $\gamma$, such that the $t$-deformed model lives on the unit circle in the $\eta$--$\gamma$ plane. That is to say, in polar coordinates, $r^2 = \eta^2 + \gamma^2 = 1$ and
$\tan\theta = \gamma/\eta = 2t/(1-t^2)$. The upper and lower half planes of the phase diagram are connected by symmetry: Reflection across the horizontal axis, $\gamma \to -\gamma$, connects the upper and lower half circles of the $t$-deformed model according to $t \to -t$; more generally, this is a transformation that swaps xy ferromagnetic correlations for xy antiferromagnetic correlations by flipping the z direction of every other spin
(and also creating an alternation in the sign of the wave function amplitudes that tracks the evenness or oddness of the area under the height field).

The purpose of this paper is to investigate and characterize the new regions at $0 < r < 1$ and $r > 1$. Note that the extended phase diagram is not everywhere frustration-free. We measure the dimer order that coexists with the Ising antiferromagnetism (z AFM) in the original \emph{t}-deformed model and show how it spreads into the broader phase space. We also identify a region  of doubly staggered Ising magnetic order (z DSM) outside the unit circle and a region of Ising ferromagnetic order (z FM) inside. 

We find that our model favors a Dyck word ground state for all points $r > 1$ in the model space, except within two small cat's-ear domes, shown in Fig.~\ref{FIG:phase-diagram-t-deformed}. We develop a representation of the Hilbert space with spins grouped into pairs whose distinct character is preserved under action by the Hamiltonian we study. To span the full Hilbert space, this representation requires that spin pairs can form conventional (xy-planar), excited (z-canted), and defect (Dyck-word spin mismatch) bonds. We establish that the  number of Dyck word defects is a good quantum number and that the ground state within the two domes is of non-Dyck form with one defect (a single mismatched pair of spins). Unlike the extended quantum critical phases in frustrated systems, which are typically bounded by a line of continuous transitions~\cite{Ramires2019,Zhao2019}, the transition here from Dyck form to non-Dyck form is a simple level crossing.

We organize this paper as follows. The model description and its full Hilbert space are discussed in Secs.~\ref{SEC:t-deformed-generalization} and~\ref{SEC:Hilbert-space}, respectively. A detailed discussion of numerical methods is presented in Sec.~\ref{SEC:t-deformed-methods}. In Sec.~\ref{SEC:t-deformed-results} we analyze the results for various types of ground states in separate subsections. Key findings of the model are summarized in Sec.~\ref{SEC:t-deformed-conclusions}. The model derivation is provided in Appendix~\ref{Append:t-deformed-model}.%@@@@@@@@@@@@@@@@@@@@@@@@@@
%@@@@@@@@@@@@@@@@@@@@@@@@@@@@@@@@
\section{Model}
\label{SEC:t-deformed-generalization}
Our starting point is the Fredkin spin chain~\cite{Salberger-RevMathPhys-16,PhysRevB-Adhikari-Beach}, a finite chain of $N$ coupled spin-half objects.
In the chain's interior, the Hamiltonian 
\begin{equation} \label{EQ:bulk_hamiltonian}
H_\text{bulk} = \sum_{i=2}^{N-1} H_i
\end{equation}
is the sum of three-site operators
\begin{equation} \label{EQ:fredkin_operator}
H_i = U_{i-1}P_{i,i+1} +
P_{i-1,i}D_{i+1}.
\end{equation}
Here, $U_i = \tfrac{1}{2}(\mathbb{1}+\sigma^z_{i})$, $D_i = \tfrac{1}{2}(\mathbb{1}-\sigma^z_{i})$, and $P_{i,i+1} = \frac{1}{4}(\mathbb{1}-\vec{\sigma}_{i}\cdot\vec{\sigma}_{i+1})$ are lone-spin-up, lone-spin-down, and spin-singlet projectors, 
with $\vec{\sigma} = (\sigma^x, \sigma^y, \sigma^z)$ denoting the Pauli matrices. 
The projector is directed at two neighboring spins, but it acts only if a third spin on the left (right) is up (down).
The boundary term $H_\text{boundary} = \alpha_1 D_1+ \alpha_N U_N$ ensures that two strong magnetic fields are applied at the chain's two open ends such that the leftmost (rightmost) spin is almost always up (down). In the numerics presented here, the external field is chosen to be $\alpha_1 = \alpha_N = \alpha = 1000$, which is the largest energy scale in the system by far. The zero-energy frustration-free ground state (GS) of the Fredkin spin chain is in the $\totsz{\text{GS}} = 0$ sector while the doubly degenerate excited states (ES) belong to $\totsz{\text{ES}} = \pm1$.

We extend the Fredkin model by replacing the singlet projector $P_{i,j}$ by a more general operator, 
\begin{equation}\begin{split}\label{EQ:generalized-projector}
\tilde{P}_{i,j}(\eta, \gamma) &= \frac{1}{4}\Bigl(\mathbb{1}-\sigma_i^z \sigma_j^z\Bigr) +\frac{\eta}{4}\Bigl(\sigma_i^z-\sigma_j^z\Bigr)\\ 
&\qquad -\frac{\gamma}{2}\Bigl( \sigma_i^+\sigma_j^- + \sigma_i^-\sigma_j^+\Bigr),
\end{split}\end{equation}
with $2\sigma^{\pm} = \sigma^x \pm i \sigma^y$ defining
the raising and lowering operators.
The two independent tuning parameters, $\eta$ and $\gamma$,
control the tendency toward Ising antiferromagnetism and the intensity
of the quantum fluctuations. The model conventions have been set so that
the $t$-deformed model coincides with $\eta^2 + \gamma^2 = 1$; the position on this circle is defined by $\eta = (1-t^2)/(1+t^2)$ and $\gamma = 2t/(1+t^2)$ [see Eq.~\eqref{EQ:t-deformed-singlet-projector} in Appendix~\ref{Append:t-deformed-model}]. 
The Fredkin point ($t=1$) is located at $(\eta = 0,\gamma = 1)$, the
``north pole'' of the unit circle.
Note that quantum fluctuations vanish along the horizontal $\gamma = 0$ line. There, the model is governed purely by the energetics, and the ground state is a single, classical configuration.

As a convenience, we transform from Cartesian coordinates to polar coordinates according to 
$\eta = r\cos\theta$ and $\gamma = r\sin\theta$. Here, $r=1$ corresponds to a $t$-deformed model with
\begin{equation}
\cos \left(\frac{\theta}{2}\right) = \frac{1}{\sqrt{1+t^2}},  \quad
 \sin \left(\frac{\theta}{2}\right) =  \frac{t}{\sqrt{1+t^2}}. 
\end{equation}
A unique point $(r=1, \theta = 90^\circ)$ represents the Fredkin model.
By construction, the ground state energy is positive inside ($r<1$), negative outside ($r>1$), and exactly zero everywhere on the unit circle. For $r \ll 1$, the model is simply an Ising ferromagnetic, largely independent of the angle $\theta$. In the other extreme limit ($r \gg 1$), the ground state is a strong function of $\theta$. At intermediate radii, there is a strong interplay between $r$ and $\theta$, and a rich phase diagram emerges.%@@@@@@@@@@@@@@@@@@@@@@@@@@@
\section{Hilbert space}
\label{SEC:Hilbert-space}
%@@@@@@@@@@@@@@@@@@@@@@@@@@@
%...............................
 \begin{figure*}
\centering
		\includegraphics[width=5.5in]{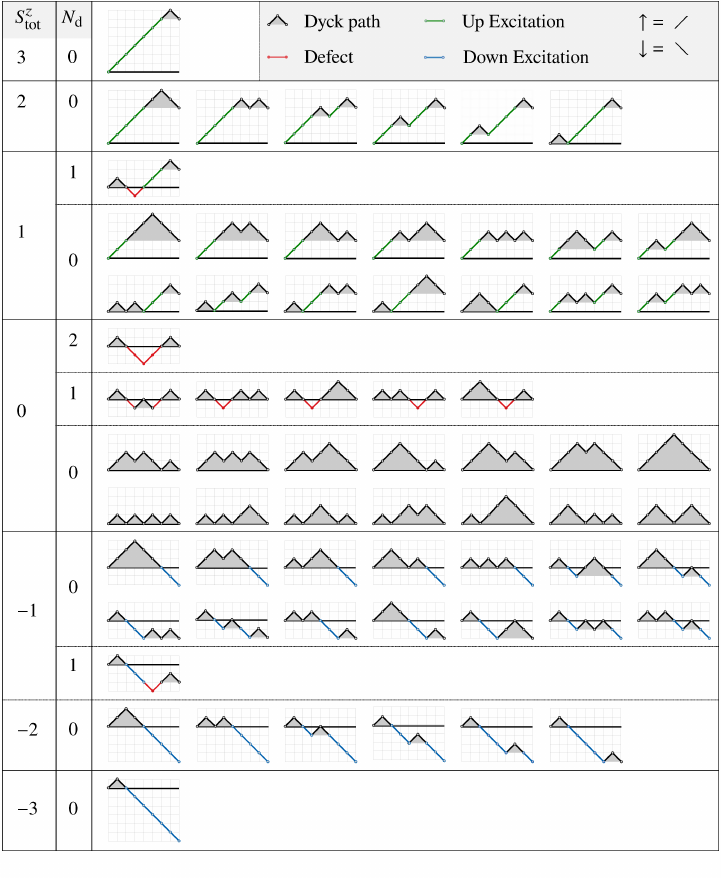}
\caption{\label{FIG:full-hilbert-space}
The full Hilbert space is depicted for the $N=8$ Fredkin
system with infinitely strong twisted boundary fields. The horizontal lines separate the subspace blocks with definite values of $\sztot$, the total spin in \emph{z}-direction, and $N_\text{d}$, the number of defects.}
\end{figure*}
%................
A Dyck path of even length $N=2n$ is a lattice path in the two-dimensional Cartesian plane with traversal from $(0,0)$ to $(n,n)$ in unit steps---either $(1,0)$ or $(0,1)$---and with the additional constraint that the path never crosses the line $y=x$. In the equivalent landscape picture, the allowed steps are the diagonals $(1,1)$ or $(1,-1)$, and the path going from $(0,0)$ to $(2n,0)$ never drops below
the horizon (the line $y=0$). The number of unique Dyck paths of length $2n$ is a sequence of numbers called the \textit{Catalan number}. The Catalan number of order $n$ is given by
\begin{equation}\label{EQ:Catalan-Number}
C_n  = \frac{1}{n+1}\binom{2n}{n} = \frac{(2n)!}{n!(n+1)!}.
\end{equation}
The Hamiltonian of the Fredkin spin chain commutes 
with $\sztot = (1/2)\sum_{i=1}^N \sigma^z_i$,
and hence the $\sztot$ is a good quantum number. 
The set of valid Dyck paths is a subspace within the $\sztot = 0$ spin sector of the Hilbert space. The full Hilbert space consists of all spin assignments with
$\sigma^z_1 = +1$, $\sigma^z_N = -1$, and $\sigma^z_i=\pm 1$ at sites $1 < i < N$. This corresponds to all random walks of the height profile starting from $h_0 = 0$ and ending at height $h_N = 2\sztot$. 

%%% TABLE I %%%
\begin{table}
	\caption{\label{TAB:bond-convention}
	         A symbol dictionary translating spin configurations into two
	         alternative representations:\ viz., height profiles; and 
	         nested and matched delimiters.}
	\begin{center}
		\begin{tabularx}{\columnwidth}{c @{\extracolsep{\fill}} ccccc}
			\hline\hline
			\makecell{Up \\ spin} &  \makecell{Down \\ spin}  &  \makecell {Dyck \\ form} & \makecell{Up \\ excitation} &  \makecell{Down \\ excitation}  &  \makecell{\\ Defect}\\
			\hline 
			$\uparrow$ &  $\downarrow$  & $\uparrow\,\downarrow$ & $ \uparrow\,\uparrow $  &  $\downarrow\,\downarrow$  &  $\downarrow\,\uparrow$\\[0.5em]
			%\hline
			$\diagup$ & $\diagdown$ & $\diagup\diagdown$ 
			& \raisebox{-0.415em}{$\diagup$}\raisebox{+0.415em}{$\diagup$} 
			& \raisebox{+0.415em}{$\diagdown$}\raisebox{-0.415em}{$\diagdown$} 
			& $\diagdown\diagup$ \\[0.5em]
			%\hline
			$\up$ & $\dn$ & $\up\dn$ & $\eo\ec$ & $\deo\dec$ &  $ \mo\mc$\\
			\hline\hline
		\end{tabularx}
	\end{center}
\end{table} 

In order to describe the full Hilbert space, we introduce two additional concepts: excitations and defects. There is some flexibility in how one defines these. In this paper, we adopt the convention illustrated in Table~\ref{TAB:bond-convention}.
Each Dyck path is in 1--1 correspondence with a Dyck word that consists of equal numbers of properly nested left and right parentheses. Although a state in Dyck form or with one or more defects can present in any spin sector, excitations only occur in the $\sztot \ne 0$ spin sectors---with up excitations only in the $\sztot > 0$ sectors and down excitations only in those with $\sztot < 0$.
To ensure a unique delimiter representation of the states, we establish the following prescription.

The population of excited bonds is
fixed within each spin projection sector 
$\sztot \in \{0, \pm1, \cdots, \pm (N/2-1) \}$.
The excitations are
\begin{equation}
 N_\text{e} = %\lvert \sztot \rvert =
    \begin{cases}
          \; \; 0, & \text{if} \; \sztot = 0,\\
      \; \; \sztot~\text{(up-spin pairs)}, & \text{if} \; \sztot > 0,\\
      \; \; \lvert \sztot \rvert~\text{(down-spin pairs)}, & \text{if} \; \sztot < 0
    \end{cases}
\end{equation}
in number
and connect (reading left to right) sites of odd and even index,
consistent with the boundary conditions.
Unlike the conventional $\up\dn$ pairs, excitations themselves cannot be nested; i.e, $\eo\ec\eo\ec$ is allowed but $\eo\eo\ec\ec$ is not.
Moreover, the representation can accommodate up to 
$N/2-1-N_e$ defects, organized in perfectly nested
form: i.e., $\mo\mo\mc\mc$ is allowed but $\mo\mc\mo\mc$ is not.
Defects occupy a position exclusively to
the left (right) of the up (down) excitations. 
Given a particular spin configuration, it is
straightforward to determine the number of defect
bonds that must appear.
We define left-cumulative and right-cumulative
height functions, 
\begin{equation}\label{EQ:cumulative-hight}
 h_i = \sum_{j=1}^i \sigma_j \ \ \text{and} \ \
 \bar{h}_i = \sum_{j=i+1}^N \sigma_j,
 \end{equation}
such that $h_0 = 0$ and $h_N = 2\sztot$ (whereas
$\bar{h}_0 = 2\sztot$ and $\bar{h}_N = 0$).
 The number of defects $N_\text{d}$ is then
\begin{equation}\label{EQ:defects}
   N_\text{d} = 
    \begin{cases}
      -\min_i h_i & \text{if} \; \sztot \ge 0,\\
      h_N - \min_i h_i = \max_i \bar{h}_i& \text{if} \; \sztot < 0.
    \end{cases}
  \end{equation} 
The remaining spins that are not participating
in an excitation or defect are grouped into
disjoint chain segments, each of which must
contain a spin configuration that is a Dyck word.
  
We have confirmed that the three-site correlated interaction 
that appears in our two-parameter tunable model 
preserves bond type, so that the number of
excitations $N_\text{e}$ and the number of defects $N_\text{d}$ are good quantum numbers. 
Hence the reshuffling of short bonds by the Hamiltonian can be used to define a canonical form for each equivalence class of states:
\begin{subequations} 
\begin{equation} \label{EQ:equiv-classes-up}
\overbrace{\!\up\dn\up\dn\cdots\up\dn\!}^{N/2-N_\text{d}-N_\text{e}-1}\,\underbrace{\!\mo\mo\cdots\mc\mc\!}_{N_\text{d}}\,\,\overbrace{\!\eo\ec\eo\ec\cdots\eo\ec\!}^{N_\text{e}}\,\up\dn
\end{equation}
or
\begin{equation} \label{EQ:equiv-classes-dn}
\up\dn\overbrace{\!\deo\dec\deo\dec\cdots\deo\dec\!}^{N_\text{e}}
\,\,\underbrace{\!\mo\mo\cdots\mc\mc\!}_{N_\text{d}}
\,\overbrace{\!\up\dn\up\dn\cdots\up\dn\!}^{N/2-N_\text{d}-N_\text{e}-1}
\end{equation}
\end{subequations}
Here, $N_\text{e}$, $N_\text{d}$,
$N/2-N_\text{d}-N_\text{e}$
represent the pair counts
of matching square-cornered brackets, angled brackets, 
and parentheses;
these take on whole number values,
limited only by the finite length of the
spin chain and by the constraints imposed
by the hard boundary conditions.
The number of spins belonging to each delimiter type,
$2N_\text{e}$ and $2N_\text{d}$
and $N-2N_\text{d}-2N_\text{e}$, is always even.
Up-canted excitations cannot reach the
right edge of the spin chain ($\ec_i : i\neq N$), and down-canted
excitations cannot reach the left ($\dec_i : i\neq 1$);
hence the lone $\up\dn$ pair on the right
and left edges in Eqs.~\eqref{EQ:equiv-classes-up}
and \eqref{EQ:equiv-classes-dn}.
Defects cannot extend to either end
($\mo_i\cdots \mc_{\!j} : i\neq 1, j\neq N$). 

The full Hilbert space for system size $N=8$ is shown in Fig.~\ref{FIG:full-hilbert-space} using the landscape representation with up ($\diagup$) and down ($\diagdown$)
slopes.
The two lowest-energy states of the two-parameter model 
defined by Eq.~\eqref{EQ:generalized-projector} live in the subspace $(\sztot, N_d) \in \{(0,0),(0,1),(\pm 1,0)\}.$ 
With the exception of the two domes, the GS is of Dyck
form and corresponds to ($\sztot = 0, N_\text{d} = 0$). 
The other important states belong to either ($\sztot = 0, N_\text{d} = 1$) or ($\sztot = \pm 1, N_\text{d} = 0$),
both of which can be represented in terms of a single defect or excitation by
\begin{equation}\label{EQ:general-form-defect-excitation}
\ket{\psi^{N_d}_{\sztot}} = \frac{1}{\sqrt{N_\text{norm}}}  \sum_{\substack{\mathcal{D}^{\prime},\mathcal{D}^{\prime \prime},\mathcal{D}^{\prime \prime \prime} \\ i, j}} g \ket{ i, j; \mathcal{D}^{\prime},\mathcal{D}^{\prime \prime},\mathcal{D}^{\prime \prime \prime}}.
\end{equation}
Here,
$\ket{ i, j; \mathcal{D}^{\prime},\mathcal{D}^{\prime \prime},\mathcal{D}^{\prime \prime \prime}}$ is a shorthand notation for the spin configuration $\ket{\mathcal{D}^{\prime}} \otimes \ket{\sigma_i} \otimes \ket{\mathcal{D}^{\prime \prime}} \otimes \ket{\sigma_j} \otimes \ket{\mathcal{D}^{\prime \prime \prime}}$,
and g = $g_{ij}(\mathcal{D}^{\prime},\mathcal{D}^{\prime \prime},\mathcal{D}^{\prime \prime \prime})$ is the corresponding wave function amplitude. The bond of is positioned at
sites $i$ and $j$ and has
($\sigma_i,\sigma_j) \in \{\: \eo\,\ec, \deo\,\dec, \mo\,\mc \:\}$ as appropriate.
The length of the Dyck words $\mathcal{D}^\prime, \mathcal{D}^{\prime \prime}$, and $\mathcal{D}^{\prime \prime \prime}$ are $i-1$, $j-i-1$, and $N-j$, respectively. The allowed values of (odd) $i$ and (even) $j$ are as follows:
\begin{equation}
\begin{aligned}
\aunderbrace[@{\,\uparrow}]{\eo}_{\,i} \cdots \aunderbrace[@{\uparrow}]{\ec}_{\!j}
&& 
i &=1,3,\ldots, N-3,\\[-0.7cm]
&& 
j &= i+1, i+3, \ldots, N-2; \\[+0.25cm]
\aunderbrace[@{\,\downarrow}]{\deo}_{\,i} \cdots \aunderbrace[@{\downarrow}]{\dec}_{\!j}
&& 
i &=3,5,\ldots, N-1,\\[-0.7cm]
&& 
j &= 
i+1, i+3, \cdots \cdots, N; \\[+0.25cm]
\aunderbrace[@{\,\downarrow}]{\mo}_{\,i} \cdots \aunderbrace[@{\uparrow}]{\mc}_{\!j}
&& 
i &=3,5,\ldots, N-3,\\[-0.7cm]
&& 
j &= i+1, i+3, \cdots \cdots, N-2.
\end{aligned}
\end{equation} 
 
 The total number of allowed configurations is
 \begin{equation}
N^\text{conf} =
 \begin{cases}
 \frac{5(n-1)(n-2)}{2(n+2)(2n-1)}C_n, &  \text{if} \; \sztot = 0, N_\text{d} = 1,\\[5pt]
 \frac{2(n-1)}{(n+2)}C_n, &  \text{if} \; \sztot = \pm 1, N_\text{d} = 0.
 \end{cases}
 \end{equation}
 For example, in the $N=8$ system with a single defect bond, 
 there are five contributing states, and Eq.~\eqref{EQ:general-form-defect-excitation} takes the form
 \begin{multline}\label{EQ:sample-example}
\ket{\psi^{1}_{0}}  = \frac{1}{\sqrt{N_\text{norm}}} \Big[ g_1\ket{\up \dn \mo \up \dn \mc \up \dn}  + g_2 \ket{\up \dn \mo \; \mc \up \dn \up \dn}  \\ + g_3 \ket{\up \dn \mo \; \mc \up \up \dn \dn} + g_4 \ket{\up \dn \up \dn \mo \; \mc \up \dn} \\ + g_5 \ket{\up \up \dn \dn \mo \; \mc \up \dn} \Big], 
\end{multline}
with normalization $N_\text{norm} = \sum_{n=1}^{N^\text{conf}} g_n^2$.
Recall that the strong external boundary fields, which
demand that $\sigma^z_1 = +1$ and $\sigma^z_N = -1$,
prevent the defect from touching either edge of the spin
chain.

%@@@@@@@@@@@@@@@@@@@@@@@@@@@
\section{Methods}
\label{SEC:t-deformed-methods}
%.........................................
\begin{figure}
\centering
		\includegraphics[width = 0.8\columnwidth]{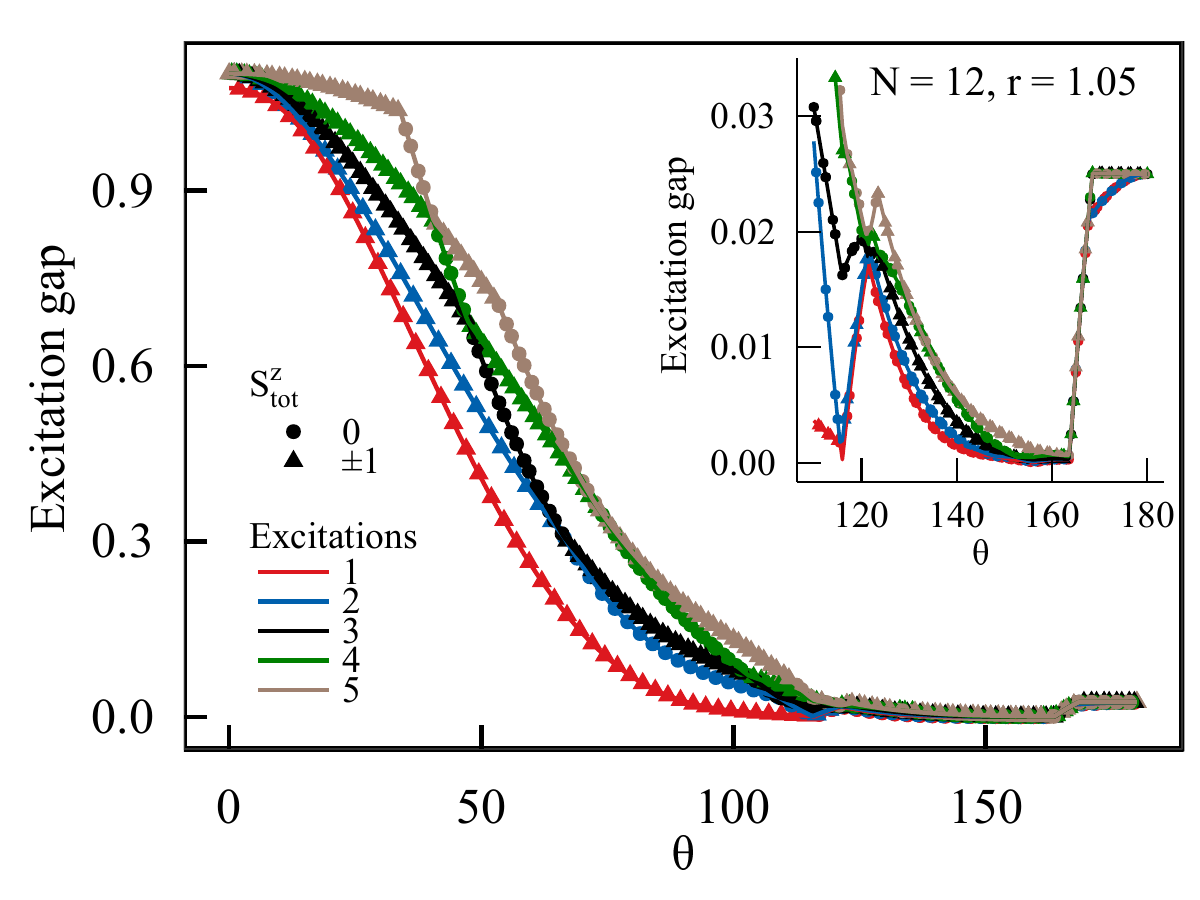}
		\caption{\label{FIG:afm-gap-spectrum-12} Exact diagonalization gap-spectrum results for  $r =1.05$ and $N=12$. Few low-lying energy spectra belong to $\lvert \sztot \rvert \le 1$ sector and it shows how difficult it is to integrate out the ground state and first excited state energy in the non-Dyck-form dome (see the north-west quadrant of Fig.~\ref{FIG:phase-diagram-t-deformed}) because the spacing between them is narrow. The first excited state (red line) make a transition from $\totsz{\text{ES}} = 1$ (solid triangle) to $\totsz{\text{ES}} = 0$ (solid circle) just before entering the non-Dyck-form ground state from the left at angle $\theta = 115.5^\circ$ as shown in the inset.}
\end{figure}
%.........................................................
Exact diagonalization (ED) is implemented as described in Ref.~\onlinecite{PhysRevB-Adhikari-Beach}. The basis-size reduction due to the discrete symmetries is not enough to significantly reduce the computational cost, so we cannot simulate large systems. To help guide our investigations, however, we have generated the full set of energy eigenstates for $N = 12$ and $N=16$ over a densely spaced mesh of ($r, \theta$) values.

To access larger sizes, we employ a DMRG algorithm implemented in the open-source C++ library ITensor~\cite{ITensor3}, taking advantage of the fact that $\sztot$ is a good quantum number in the model. We are mindful of the fact that the high level of entanglement in the vicinity of the Fredkin model and its mirror point at the ``south pole'' ($r \ge 1$ and $90^\circ \le \theta \lesssim 110^\circ$ or $250^\circ \lesssim \theta \le 270^\circ$) requires us to keep many states; farther away, we can be more cavalier about truncating the DMRG basis set. We are also careful about issues of convergence: in and around the two domes of the phase diagram (roughly corresponding to a region $1 \le r \lesssim 1.2$ and $110^\circ \lesssim \theta \lesssim 260^\circ$) 
an unexpectedly large number of sweeps is required, because the low-lying energy levels are very closely spaced (see Fig.~\ref{FIG:afm-gap-spectrum-12}). 

Moreover, the convergence is strongly biased by the choice of initial trial state, because the nature of the low-lying states sometimes changes abruptly with a small change in $r$ or $\theta$ values. As a workaround, we make a list of possible low energy configurations using ED results for $N \le 16$. Comparable configurations are then used to seed the DMRG calculations for bigger system sizes. The first excited state belongs to the $\totsz{ES} = 0$ spin sector in most parts of this region. So, we calculated the two orthogonal states in $\sztot = 0$ having the lowest eigenvalues using many possible trial states. Then, the ground state and the first excited state are found by sorting two lowest energies in the $\sztot = 0$ spin sector and the lowest energy in the $\sztot =1$ spin sector. 

We have employed a very conservative convergence criterion: the DMRG algorithm runs through $10N$ sweeps using an adaptive truncation cutoff at relative error $10^{-12}$ with maximum bond dimension $15N$. We have benchmarked our DMRG results to near double-precision floating-point accuracy against ED results for $N \leq 16$.
The DMRG computation is carried out for all lattice sizes that are multiples of 4 up to $N=60$, over a tight mesh of tuning parameter values $r = 0,0.025, \ldots, 2.975,3$ and $\theta = 0^\circ,1^\circ, \ldots, 359^\circ, 360^\circ$. The most expensive of those simulations corresponds to $600$ sweeps with a maximum bond dimension of 900. 
Various physical quantities are computed in the ground state as a function of the tuning parameters $r,\theta$; i.e., $O(r,\theta) = \expectation{\hat{O}} = \bra{\psi_0(r,\theta) } \hat{O} \ket{\psi_0(r,\theta)}$. These include the spin profile $\expectation{\sigma_i^z}$, dimer profile $\expectation{\sigma_i^z \sigma_{i+1}^z} -  \expectation{\sigma_j^z}\expectation{\sigma_{j+1}^z}$, dimer order parameter
\begin{alignat}{2}
\expectation{d_{\parallel}} &= \frac{1}{N}\sum_{j=1}^{N-1}c_j \bigl[ \expectation{\sigma_j^z \sigma_{j+1}^z} - \expectation{\sigma_j^z}\expectation{\sigma_{j+1}^z} \bigr]
\end{alignat}
where
\begin{equation*}
    c_j = (-1)^j
    \begin{cases}
      1/2, & \text{if  $j = 1$ or $N$} \\
      1, & \text{otherwise}
    \end{cases}
  \end{equation*}
Ising ferromagnetic order parameter (z FM) 
\begin{alignat}{2}
\expectation{m_{\parallel}(\text{FM})} &= \frac{1}{N}\sum_{j=1}^N \expectation{\sigma_j^z},
\end{alignat}
Ising antiferromagnetic order parameter (z AFM) 
\begin{alignat}{2}
\expectation{m_{\parallel}(\text{AFM})} &= \frac{1}{N}\sum_{j=1}^N (-1)^{j}\expectation{\sigma_j^z},
\end{alignat}
doubly staggered Ising magnetic order parameter (z DSM) 
\begin{alignat}{2}
\expectation{m_{\parallel}(\text{DSM})} &= \frac{1}{N}\sum_{j=1}^N c_j \expectation{\sigma_j^z}
\end{alignat}
where
\begin{equation*}
    c_j =
    \begin{cases}
      +1, & \text{if $j= 0,1\ (\text{mod}\,4)$}\\
      -1, & \text{if $j= 2,3\ (\text{mod}\,4)$}
    \end{cases}
  \end{equation*}
 the xy-plane ferromagnetic order parameter (xy FM) 
\begin{equation}
\expectation{m_{\perp}^2(\text{xy FM})} = \frac{1}{N^2}\sum_{i,j=1}^N \expectation{\sigma_i^+\sigma_j^-+\sigma_i^-\sigma_j^+},
\end{equation}
and the xy-plane antiferromagnetic order parameter (xy AFM) 
\begin{equation}
\expectation{m_{\perp}^2(\text{xy AFM})} = \frac{1}{N^2}\sum_{i,j=1}^N (-1)^{i+j}\expectation{\sigma_i^+\sigma_j^-+\sigma_i^-\sigma_j^+}.
\end{equation}
%@@@@@@@@@@@@@@@@@@@@@@@@@@@@@
\section{Results and discussion}
\label{SEC:t-deformed-results}
Unlike the upper half-plane where wave function amplitudes are all positive, the wave function amplitudes in the lower circular plane contain an admixture of positive and negative signs. About the horizontal line ($\gamma = 0$), diagonal ($\sigma^z$ dependent only) and off-diagonal (products of $\sigma^+$ and $\sigma^-$) measurements are symmetric and antisymmetric, respectively. For example, xy FM and xy AFM measurements in Fig.~\ref{FIG:phases-60}(b) have peaks at $(r, \theta) = (1, 90^\circ)$ and  $(1,270^\circ)$, respectively. So, we discuss our diagonal measurement results  only for the upper semicircular plane, and a similar explanation applies to the lower semicircular plane. We analyze the ground state properties of the system at distinct regions of the phase space separately in the Secs.~\ref{SUB-SEC:classical-result}--\ref{SUB-SEC:Non-Dyck-Form}. 
%...........................................................
\subsection{Fluctuation-free limit ($\gamma = 0)$}\label{SUB-SEC:classical-result}
The classical energy of Eq.~\eqref{EQ:fredkin_operator} is given by 
\begin{equation}\label{EQ:classical-energy}
    E(\eta, \gamma = 0)_i =
    \begin{cases}
      \; \; (1+\eta)/2, & \text{if $\ket{ \upquiet \ud}$  or $\ket{ \ud \dnquiet}$}\\
      \; \; (1-\eta)/2, & \text{if $\ket{ \ud \upquiet}$  or $\ket{\dnquiet \ud}$}\\
      \; \; 0,  & \text{otherwise}
    \end{cases}
  \end{equation}
The energy difference ($\Delta E _ i  = \pm \eta$) in Eq.~\eqref{EQ:classical-energy} can be viewed as a movement of a short bond $\ud$ to the left or right; i.e., $\upquiet \ud \Leftrightarrow \ud \upquiet$ or $ \ud \dnquiet \Leftrightarrow \dnquiet \ud$. In Table~\ref{TAB:config-and-energy}, we summarize a list of several low-lying energy configurations of the classical model using Eq.~\eqref{EQ:classical-energy}.
%...........................

%%% TABLE II %%%
\begin{table}
	\caption{\label{TAB:config-and-energy}
	         A list of relevant spin configurations for the two low-lying
	         states of the classical model of size $N =8$. States $\ket{8}$
	         and $\ket{9}$ are excluded from the Hilbert space in the limit
	         ($\alpha \to \infty$) of infinitely strong boundary fields.}
	\begin{center}
		\begin{tabularx}{\columnwidth}{l @{\extracolsep{\fill}} cc}
			\hline\hline
			State & Configurations & $E(N,\eta, \gamma = 0, h)$  \\
			\hline
			$\ket{1}$ (z FM) & $\ket{\eo \ec \eo \ec \eo \ec \ud}$ & $(1+\eta)/2$\\
			                 & $\ket{\ud \deo\dec\deo\dec\deo\dec}$\\[0.15cm]
			%\hline
			$\ket{2}$        & $\ket{\ud \eo \ec \eo \ec \ud}$     & 1\\
			                 & $\ket{\ud \deo\dec\deo\dec \ud}$\\[0.15cm]
			%\hline
			$\ket{3}$        & $\ket{\ud \ud \ud \ud}$             & $(N-2)(1-\eta)/2$\\[0.15cm]
			%\hline
			$ \ket{4}$       & $\ket{\eo \ec \ud \up \dn \up \dn}$ & $(N-4)(1-\eta)/2 + (1+\eta)/2 $\\[0.15cm]
			%\hline
			$\ket{5}$ (DW)   & $\ket{\up \up \up \ud \dn \dn \dn}$ & $1+\eta$\\[0.15cm]
			%\hline
			$\ket{6}$ (z DSM)& $\ket{\up \ud \dn \up \ud \dn}$     & $N(1+\eta)/4$, $N = 0\ (\text{mod}\,4)$\\[0.15cm]
			%\hline
			$\ket{7}$        & $\ket{\ud \mo \mc \up \ud \dn}$     & $(N-2)(1+\eta)/4$, $N = 0\ (\text{mod}\,4)$\\
			                 & $\ket{\up  \ud  \dn \eo \ec  \ud}$ \\
      		                 & and others\\[0.15cm]
			%\hline
			$\ket{8}$        & $\ket{\ud \ud \ud \eo \ec}$         & $(N-3)(1-\eta)/2+\alpha$\\
			                 & $\ket{\deo \dec \ud \ud \ud}$\\[0.15cm]
			%\hline
			$\ket{9}$        & $\ket{\mo \ud \ud \ud \mc}$         & $(N-2)(1-\eta)/2+2\alpha$\\
			\hline\hline
		\end{tabularx}
	\end{center}
\end{table}

 The lowest two energy configurations of this classical model are discussed in Secs.~\ref{SubSec:positive-x-axis} and~\ref{SubSec:negative-x-axis}.
%...........................................................
\subsubsection*{\emph{On the positive x-axis ($\theta = 0, \eta = r$)}}
\label{SubSec:positive-x-axis}

%%% TABLE III %%%
\begin{table}
	\caption{\label{TABLE:GS-ES-gap-0}
             The table summarizes the ground state, first excited state, and 
             the excitation gap at $\theta = 0^\circ$. The corresponding spin
             configuration of the state (GS or ES) are shown in
             Table~\ref{TAB:config-and-energy}.}
	\begin{center}
		\begin{tabularx}{\columnwidth}{c @{\extracolsep{\fill}} ccc} 
			\hline \hline
			Radial coordinate & $\ket{\text{GS}} $  & $\ket{\text{ES}} $ & $\Delta(N,r,\alpha)$\\
			\hline
			$r=0$                                & $\ket{1}$ & $\ket{2}, \ket{5}$,  & $1/2$\\
			                                     &           & and others\\[0.15cm]
			%\hline
			$0 < r < r_\text{c}(N)$              & $\ket{1}$ & $\ket{2}$            & $(1-r)/2$\\[0.15cm]
			%\hline
			$r_\text{c}(N) < r < r_\text{cc}(N)$ & $\ket{1}$ & $\ket{3}$            &  $(N-1)(1-r)/2-1$\\[0.15cm]
			%\hline
			$r_\text{cc}(N) < r < 1$             & $\ket{3}$ & $\ket{1}$            & $1-(N-1)(1-r)/2$\\[0.15cm]
			%\hline 
			$r=1$                                & $\ket{3}$ & $\ket{4}$            & $r+(r-1)/2$\\
			& & and others\\[0.15cm]
			%\hline 
			$1 < r < \alpha$                     & $\ket{3}$ & $\ket{4}$            & $r+(r-1)/2$\\[0.15cm]
			%\hline
			$r = \alpha $                        & $\ket{3}$ & $\ket{4}$, $\ket{8}$ & $\alpha+(r-1)/2$\\[0.15cm]
			%\hline
			$\alpha < r < 2\alpha+1$             & $\ket{3}$ & $\ket{8}$            & $\alpha+(r-1)/2$\\[0.15cm]
			%\hline
			$r =  2\alpha+1$                     & $\ket{3}$ & $\ket{8}$, $\ket{9}$ & $2\alpha$\\[0.15cm]
			%\hline
			$r >  2\alpha+1$                     & $\ket{3}$ & $\ket{9}$            & $2\alpha$\\
			\hline\hline
		\end{tabularx}
	\end{center}
\end{table}

 The ground state, excited state, and the excitation gap are shown in Table~\ref{TABLE:GS-ES-gap-0}. Let us define two critical radii $r_\text{c}(N) = (N-4)/(N-2)$ and $r_\text{cc}(N) = (N-3)/(N-1)$. At $r=0$, only the first term survives in Eq.~\eqref{EQ:generalized-projector}. The ground state favors the doubly degenerate z FM state and the first excited state belongs to the highly degenerate states of types $\ket{2}$ and $\ket{5}$ with an excitation gap $\Delta = 1/2$. For a finite system size \textit{N}, the first-order phase transition from z FM to z AFM occurs exactly at $r_\text{cc}(N)$. In the thermodynamic limit, the phase transition occurs exactly at $r_\text{cc}(N \to \infty) = 1$. Both z FM and z AFM are fully polarized at this angle and they belong to the $\totsz{GS} = (N-2)/2$ and $\totsz{GS} = 0$, respectively. For $r \ge 1$, the ground state favors z AFM. The excitation gap is $\Delta \text{(z FM)} = (1-r)/2$ for $r <1$, and $\Delta \text{(z AFM)} = r+(r-1)/2$ for $r \ge 1$ in the limit $\alpha \to \infty$. 
  
 The ground state is independent of the field strength $\alpha$ (tested for $\alpha \ge 1$). However, the magnitude of field strength affects the measurement of the excitation gap if $r \ge \alpha$ and the angle $\theta$ is small. The spin-flip at the boundary adds extra energy $\alpha$ to the system resulting $\totsz{\text{ES}}=0$ excitation in stead of $\totsz{\text{ES}}=1$. At exactly $r=\alpha$, the state $\ket{8}$ with the excitation gap $\Delta(r, \alpha) = \alpha+(r-1)/2$ is equal to the state $\ket{4}$. So, the state $\ket{8}$ is the excited state in the range $\alpha < r < 2\alpha+1$. At exactly $r=2\alpha+1$, the state  $\ket{9}$ with two spins flips at the boundary overlap with previous doubly degenerate excited states giving common gap $\Delta(\alpha) = 2\alpha$. For $r>\alpha$, a unique state $\ket{9}$ with a constant gap $\Delta(\alpha) = 2\alpha$ is the excited state.
 %.............................................................................................
\subsubsection*{\emph{On the negative x-axis ($\theta = 180^\circ$, $\eta = -r$)}}
\label{SubSec:negative-x-axis}

%%% TABLE IV %%%  
\begin{table}
	\caption{\label{TABLE:GS-ES-gap-180}
	         The table summarizes the ground state, excited state, and the 
	         excitation gap at $\theta = 180^\circ$. The corresponding spin 
	         configurations of the state (GS or ES) are shown in 
	         Table~\ref{TAB:config-and-energy}.}
	\begin{center}
		\begin{tabularx}{\columnwidth}{c @{\extracolsep{\fill}} ccc} 
			\hline\hline
			Radial coordinate & $\ket{\text{GS}} $    & $\ket{\text{ES}} $ &  $\Delta(N,r)$\\
			\hline
			$r=0$             & $\ket{1}$             & $\ket{2}$, $\ket{5}$, & $1/2$\\
			                  &                       & and others\\[0.15cm]
			%\hline
			$0 < r < 1$       & $\ket{1}$             & $\ket{5},$             & $(1-r)/2$\\
			                  &                       & $\ket{\ud  \mo \mo \mc \mc \ud}$, \\
			                  &                       & $\ket{\eo \ec \up \up \ud \dn \dn}$, \\
			                  &                       & and others\\[0.15cm]
			%\hline
			$r=1$             & $\ket{5}$             & $\ket{2}$,              & $1$ \\
			                  &                       & $\ket{\eo  \ud  \ec  \eo \ec \ud}$, \\
			                  &                       & $\ket{\up \dn \up  \up \up \dn  \dn \dn}$, \\
			                  &                       & and others\\[0.15cm]
			%\hline 
			$ 1 < r < \infty$ & $\ket{6}$  & $\ket{7}$  & $(r-1)/2$ \\
			\hline\hline
		\end{tabularx}
	\end{center}
\end{table}

The ground state, excited state, and the excitation gap are shown in Table~\ref{TABLE:GS-ES-gap-180}. For $0 <  r < 1$, the ground state is the doubly degenerate z FM, and the first excited state is highly degenerate. At $r=1$, the z AFM is highly penalized, but all other states including the domain wall (D-Wall) have the same energy resulting in highly degenerate ground states with $\totsz{GS} = 0, \pm 1, \cdots,  \pm (N-2)/2$. The highly degenerate excited state belongs to different sectors $S^z_\text{tot} = 0, \pm 1, \cdots, \pm (N-4)/2$. For $1 < r < \infty$ and  $N = 0\ (\text{mod}\,4)$, the ground favors a state that forms a repeated patterns of four spins, z DSM. The degenerate excited states belong to $\totsz{ES} = 0, 1$ sectors. The excitation gap is independent of the system size. The $N = 2\ (\text{mod}\,4)$ sizes are excluded from this work to avoid ambiguity because their ground state is degenerate in the range $1 < r < \infty$. For example, $E(N,\eta) = (N-2)(1+\eta)/4$ for $ \ket{ \ud  \mo \mc \ud}$, $\ket{ \eo \ec \up  \ud \dn }$, and many other states.
%.................................................................................
\subsection{Unit circle ($r=1$)}
\label{SUB-SEC:r-equal-1}
On the unit circle, the ground state is the area-weighted sum of the Dyck-form. The zero-energy unique ground state belongs to $\totsz{GS} = 0$, but the excitations are doubly degenerate in $\totsz{ES} = \pm 1$ spin sectors.  The left unit semicircle is featureless (no order), and the excitation gap vanishes exponentially fast.
On the right unit semicircle, the ground state is ordered, and the excitations are gapped. The spectral gap obeys the threshold criteria for frustration-free spin systems with boundary~\cite{Lemm-gaps-JMP-2019}. The peaks of z AFM ($= \cos \theta/2$) and the excitation gap at ($r,\theta) = (1, 0^\circ$) both vanish smoothly at the Fredkin point ($1, 90^\circ$). There is a strong dimer order at $(1,\approx 65^\circ$) that gradually weakens with changing angle until it disappears completely at ($1, 0^\circ$) and ($1, 90^\circ$). For finite system sizes, the xy FM is smeared out in the vicinity of the Fredkin point, but it collapses to a delta function in the thermodynamic limit. Although diagonal measurements are symmetric about the horizontal line ($\gamma = 0$), off-diagonal measurements are antisymmetric. So, strong xy AFM is measured at the bottom of the unit circle that was not included in the origin $t$-deformed model.

\subsection{Inside the unit circle ($r<1$)}\label{SUB-SEC:r-less-than-1}
\begin{figure}
\centering
	\includegraphics[width=0.45\columnwidth]{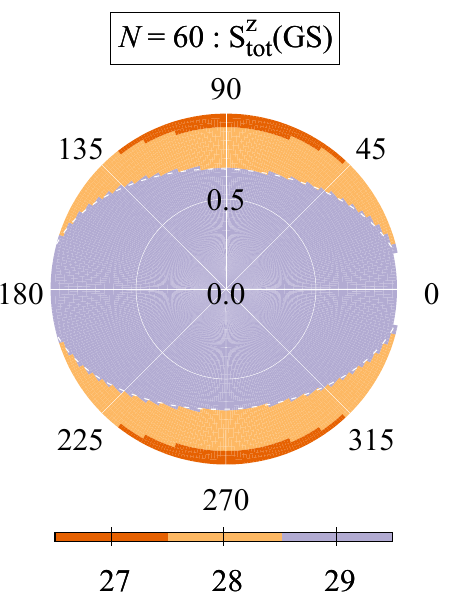}\hspace{0.025\columnwidth}
		\includegraphics[width=0.45\columnwidth]{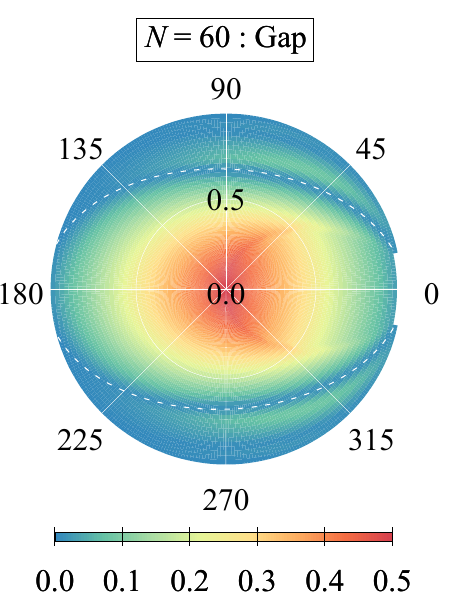}
		\caption{\label{FIG:fm-phase} These are discrete colormaps for $N=60$ system in the parameter regime $r<1$. The interior labels mark the radii, and the outer labels denote the angles in degrees.  Left: Three different colors denote the ground state spin sectors $ \totsz{GS} = (N-2)/2$, $(N-4)/2$, and $(N-6)/2$ for the innermost, middle, and outermost elliptical regions, respectively. The ferromagnetic order weakens along the vertical diagonal away from the center. Right: The excitation gap is exactly $\Delta = 1/2$ at the center and vanishes as $r \to 1$. The gapless elliptical boundaries coincide with the spin sector crossover in the ground state.}	
\end{figure}

For $\theta = 180^\circ$, the phase transition from z FM to D-Wall state occurs exactly at $r=1$, independent of the system size. However, for $\theta = 0$, the phase transition from z FM to z AFM occurs exactly at $r_\text{cc}(N)$ for a finite system size, \textit{N}. For  $r_\text{cc}(N) < r < 1$, the true nature of the system in the ground state is suppressed by the finite size effect where the ground state favors z AFM only for the finite system.
Although the ground state properties are independent of system size for $r \le r_\text{cc}(N)$, the excitation gap does only for $r \le r_\text{c}(N)$. In the thermodynamic limit, $r_\text{c}(\infty) \to  r_\text{cc}(\infty) \to 1$. In Fig.~\ref{FIG:fm-phase}, we omitted the data in the vicinity of ($r \lesssim 1, \theta \approx 0$) to exclude the finite-size effect.  Along the vertical line ($\eta = 0$) in the phase diagram, the ground state belongs to $\totsz{GS} = (N-2)/2$ sectors for $0 \le r  \lesssim 2/3$.  As $r$ increases, the spin sector decreases gradually first and then exponentially fast as ($r \lesssim 1, \theta = 90^\circ$) resulting in $\totsz{GS} = 0$ at the Fredkin point. 
The model is gapped at the center, $\Delta(r=0) = 1/2$, and gapless along the unit circle.  The ellipses with gapless boundary correspond to the ground state level crossing in the spin sectors $\totsz{GS}$: $(N-2)/2 \to  (N-4)/2, (N-4)/2 \to  (N-6)/2$, and so on as shown in Fig.~\ref{FIG:fm-phase}. The ground state and the excited state belong to different spin sectors and satisfy $\totsz{GS} = \totsz{ES} \pm1$ ($-$ in the transition window followed by the gapless boundary and $+$ elsewhere). There is no quantum fluctuation the in $\totsz{GS} = (N-2)/2$ sector because promoting the $N^\text{th}$ down spin to up costs an additional energy $h$ to the system. So, the ground state energy is $E_0 = (1+\eta)/2$ (independent of $\gamma$) with the unique z FM inside the innermost ellipse shown in Fig.~\ref{FIG:fm-phase}.

\subsection{Outside the unit circle ($r>1$)}\label{SUB-SEC:r-greater-than-1}
\begin{figure}
\centering
	\includegraphics[width=0.45\columnwidth]{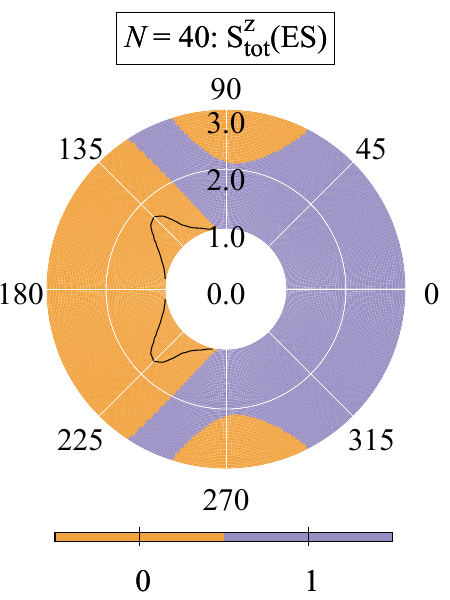}
	\hspace{0.025\columnwidth}
		\includegraphics[width=0.45\columnwidth]{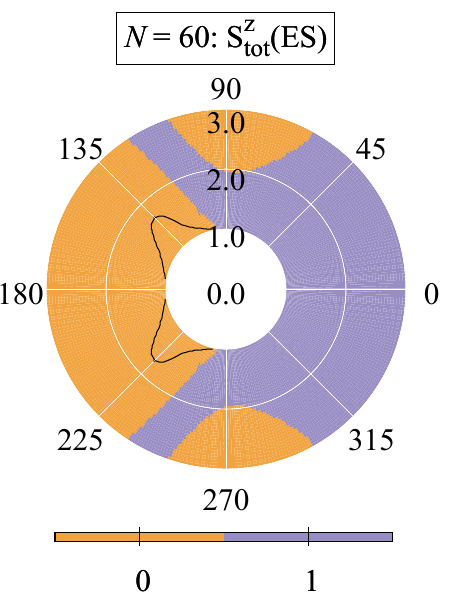}
	\includegraphics[width=0.45\columnwidth]{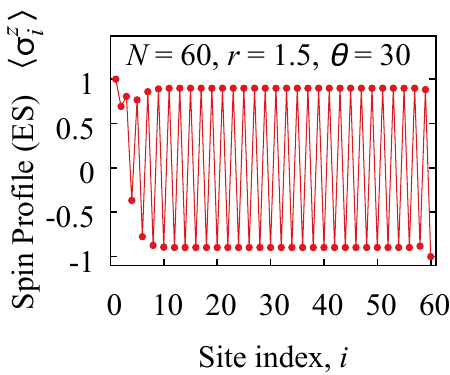} \hspace{0.025\columnwidth}
		\includegraphics[width=0.45\columnwidth]{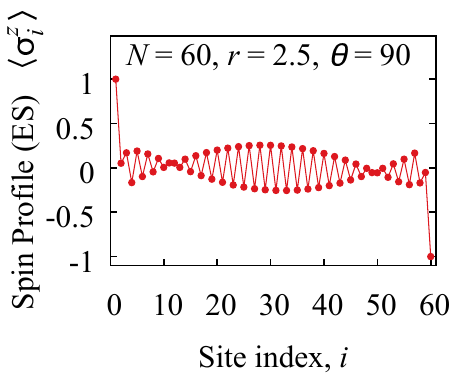}
	\caption{\label{FIG:QN1}Top: The colormap denotes the discrete values of $\totsz{\text{ES}}$, the total spin projection of the first excited state. It also shows the inward drift of some boundaries (viz., the orange zones near $90^\circ$ and $270^\circ$) with increasing system size. Bottom: Spin profiles measured in the excited state for system size $N=60$ at parameter values $(r=1.5,\theta = 30^\circ)$ and $(r=2.5,\theta = 90^\circ)$. These profiles correspond to excitations of distinct character.}	
\end{figure}

\begin{figure*}
\centering
	\includegraphics[width=0.3\textwidth]{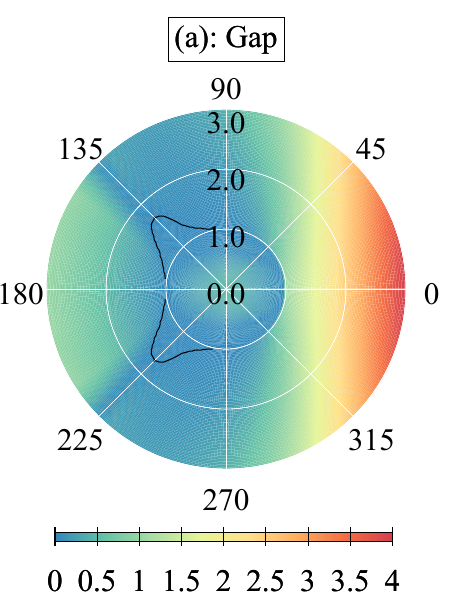}\hspace{0.25cm}
		\includegraphics[width=0.3\textwidth]{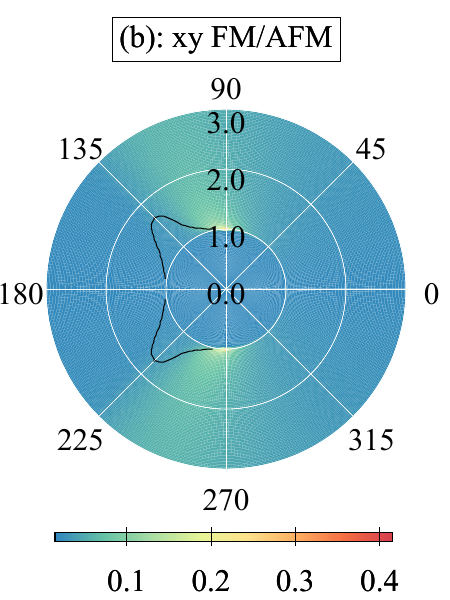}\hspace{0.25cm}
			\includegraphics[width=0.3\textwidth]{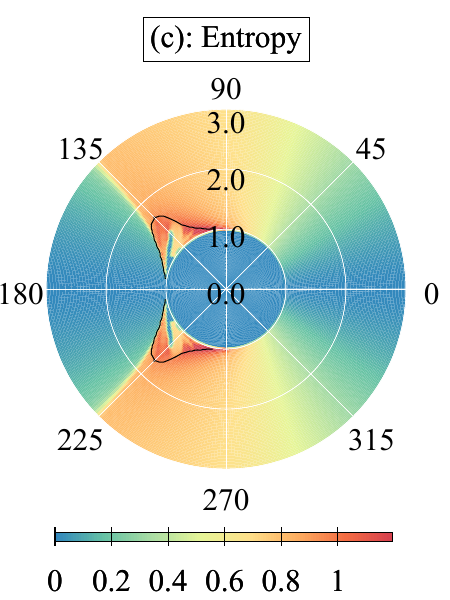}\vspace{0.5cm}
				\includegraphics[width=0.3\textwidth]{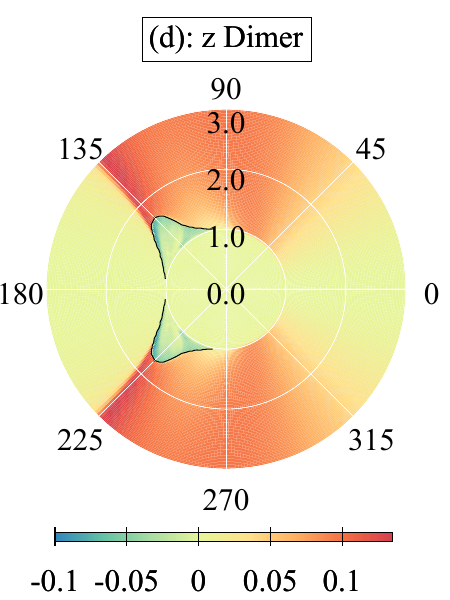}\hspace{0.25cm}
					\includegraphics[width=0.3\textwidth]{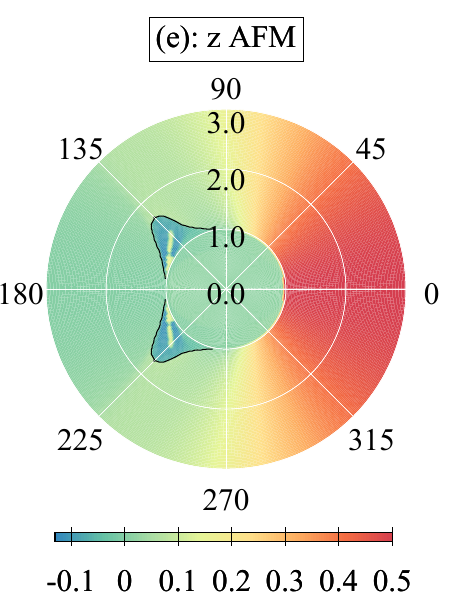}\hspace{0.25cm}
						\includegraphics[width=0.3\textwidth]{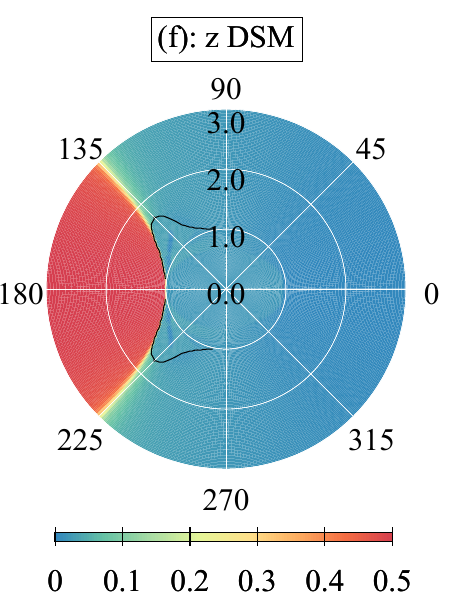}
	\caption{\label{FIG:phases-60}Sample color maps for the finite system $N=60$ where black solid line represent the domes boundaries. (a) Excitation gap. (b) The xy-directed FM (AFM) on the upper (lower) semicircular regions. (c) The von Neumann entanglement entropy. (d) The dimer order parameter showing slightly varying magnitudes at different parts of the phase diagram. (e) The Ising antiferromagnetic order parameter, z AFM. (f) Doubly staggered Ising magnetic order parameter, z DSM.}
\end{figure*}

The ground state is Dyck-form everywhere except in two Cat ears like plane domes residing on the left semicircular plane of the phase space (see Figs.~\ref{FIG:QN1} and \ref{FIG:phases-60}). The non-Dyck-form and Dyck-form ground states are discussed in Secs.~\ref{SUB-SEC:Non-Dyck-Form} and ~\ref{SUB-SEC:Dyck-Form}, respectively.
%........................................................................................................................................
%................................
%........................................................................................................................................
\begin{figure}
\centering	
	\includegraphics[width=0.45\columnwidth]{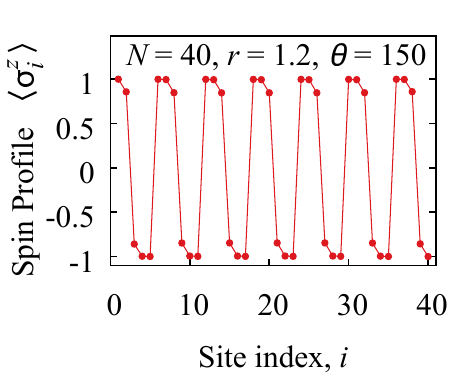}\hspace{0.025\columnwidth}
	\includegraphics[width=0.45\columnwidth]{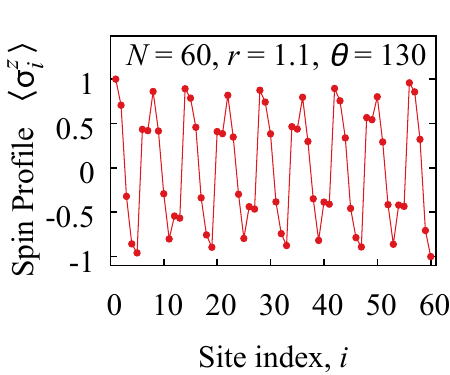}\vspace{0.025\columnwidth}
	\includegraphics[width=0.45\columnwidth]{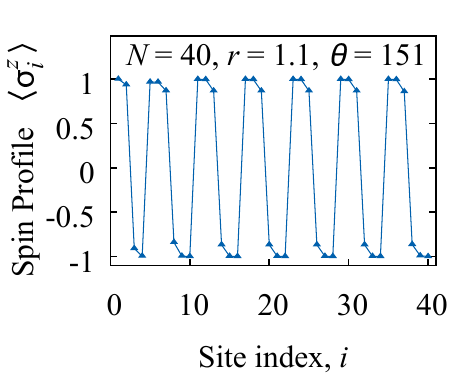}\hspace{0.025\columnwidth}
	\includegraphics[width=0.45\columnwidth]{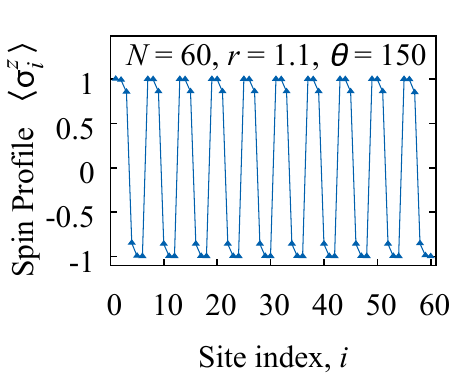}
	\caption{\label{FIG:spin-profile-dome}Top row: The two spin profiles shown (red circles) are typical of the non-Dyck-form ground states that arise in the dome. The individual configurations $\up \up \dn \dn \mo \up \up \up \dn \dn \dn \cdots \up \up \up \dn \dn \dn \mc \up \up \dn \dn $  and $\up \up \dn \dn \mo \up \up \up \up \dn \dn \dn \dn \cdots \up \up \up \up \dn \dn \dn \dn \mc \up \up \dn \dn $ dominate the expectation values. The red angled brackets in these spin configurations denote a single defect. Bottom row: For some system sizes, a narrow band of Dyck-compatible ground states traces out a line across the phase space inside the dome. There, the spin profile (blue triangles) is defect-free. The dominant contributions to measurements in the bottom left and bottom right panels are $\up \up \dn \dn \up \up \up \dn \dn \dn \cdots \up \up \up \dn \dn \dn$ and $ \up \up \up \dn \dn \dn \cdots \up \up \up \dn \dn \dn$, respectively.}	
\end{figure}
%.........................................................
\begin{figure}
\centering
\includegraphics[width=0.45\columnwidth]{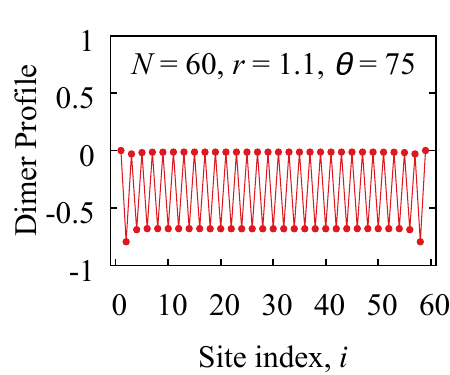} \hspace{0.025\columnwidth}
	\includegraphics[width=0.45\columnwidth]{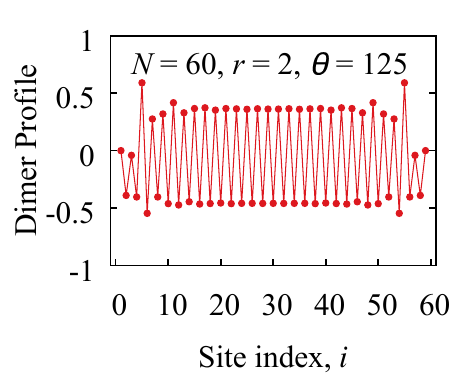}
	\caption{\label{FIG:dimer-profile-60}The dimer profile measured in the ground state for size $N = 60$ at two points in the model-parameter space, $(r=1.1,\theta = 75^\circ)$ and $(r = 2,\theta = 125^\circ)$. The data show two distinct dimer patterns.}
\end{figure}
%.................................................................................................
\subsubsection*{\emph{Non-Dyck-form ground state}}\label{SUB-SEC:Non-Dyck-Form}
The non-Dyck-form unique ground state is observed for $N \ge 10$ in the dome residing its base along the unit circle and peaked at $(r \approx 1.7, \theta \approx135^\circ)$ as shown in Fig.~\ref{FIG:phases-60}. The ground state shows idiosyncratic nature while making the phase transition from z AFM at angle $0^\circ$  through z DSM order at angle $180^\circ$ by creating different numbers of peaks in the spin profile as shown in Fig.~\ref{FIG:spin-profile-dome} (Top). The nature of forming a group of spins cluster and transition between them make the system strongly size-dependent non-Dyck-form ground state. The Dyck-form and non-Dyck-form do not co-exist in a particular state because they have distinct quantum numbers (see Sec.~\ref{SEC:Hilbert-space}). In two domes, the ground state favors $\sztot = 0$ and $N_\text{d} = 1$. The first excited state heavily depends on ($r,\theta$) values and belongs to $(\sztot, N_\text{d}) \in \{(0,0),(0,1) \}$ subspace of the Hilbert space.

Since the states are not entangled, DMRG should work perfectly but this is not the case. The small energy scale (see Fig.~\ref{FIG:afm-gap-spectrum-12}) hinders to find the global minima, and the solution is also biased to the initial trial state (noise observed in the domes). In some parts of the non-Dyck-form ground state, the lowest-lying energy states are almost continuum where the Dyck-form ground state appears for few selected system sizes and special tuning points as a coincidence [spin profile shown in Fig.~\ref{FIG:spin-profile-dome} (bottom) and line observed in Fig.~\ref{FIG:phases-60} (c and e) in the domes].  Additionally, the strong size-dependent nature observed in the measured values makes extrapolation almost impossible (see Fig.~\ref{FIG:afm-gap-spectrum-12}).  We believe that this non-Dyck-form dome appears because of frustration arising from many competing phases surrounding it. The size of the dome is robust and already conversed for $N=60$ as tested against $N=120$.
%...........................................................................................
\subsubsection*{\emph{Dyck-form ground state}}\label{SUB-SEC:Dyck-Form}
The unique ground state is in $\totsz{\text{GS}} = 0$ sector everywhere, and the first excitation belongs to $\totsz{ES} = 0, \pm1$ as shown in Fig.~\ref{FIG:QN1} (Top).
The hyperbolic regions containing Dyck-form ($\totsz{ES} = 0, N_\text{d} =0$) excitation emerges for $N\ge10$ and broaden with the increasing system size. As a result, the vertex approaches ($r \approx 1.7, \theta \lesssim 90^\circ$; peak of the non-Dyck-form dome) in the thermodynamic limit.
 Unlike $\totsz{\text{ES}} = 1$ excitation where the spin-flip occurs at one end of the chain, even the number of spins flip resulting in the Dyck-form excitation as shown in Fig.~\ref{FIG:QN1} (Bottom).

The excitation is gapless only along the unit circle and the boundary of non-Dyck-form, leaving the system gapped everywhere on the phase space. The excitation gap increases along the radial direction away from the center for all angles. In the angular direction, the gap maximum at $\theta = 0^\circ$ decreases gradually and attains its minima  at the hyperbolic boundary on the left semicircular region. In z DSM, the region enclosed by the hyperbolic boundary, the excitation gap does not vary much with the change in angle $\theta$.

The xy-directed ferromagnetic correlation observed at the Fredkin point appears to leak away from the center on the left side of the vertical line as shown in Fig.~\ref{FIG:phases-60} (b). Although von Neumann entropy shows a significant boost in that region, it does not scale with the system size because the system is gapped and follows the area law of entanglement entropy. Unfortunately, this apparent entangled state requires a larger bond dimension in DMRG calculation resulting in additional time complexity.

The fully saturated z AFM at $\theta = 0^\circ$ decreases continuously in the angular direction until it completely vanishes at the z DSM boundary as shown in Fig.~\ref{FIG:phases-60}(e). In the radial direction away from the center, z AFM first increases gradually,  and then saturates to a finite value.
The dimer order is absent in both z DSM and strong z AFM regions. The dimer order peak at $(r=1,\theta \approx 65^\circ)$ shifts toward the left part of the phase space with the increasing value of radius $r$ as can be seen in Fig.~\ref{FIG:phases-60}(d). 
The slightly different magnitude of dimer order observed in Fig.~\ref{FIG:phases-60}(d) are shown by the distinct dimer profiles in Fig.~\ref{FIG:dimer-profile-60}.
%@@@@@@@@@@@@@@@@@@@@@@@@@@@@@@@
\section{Conclusions}\label{Conclusions}
\label{SEC:t-deformed-conclusions}

In Sec.~\ref{SEC:Hilbert-space}, we discussed how to represent the Hilbert space using the language of matching and nested spin pairs. Conventional pairs, with a spin up to the left and spin down to the right, are the building blocks of the Dyck word ground state. By promoting certain conventional bonds to a bond of different character---either an excited bond that cants the spin state out of the xy plane or a defect bond that carries a Dyck word mismatch---we were able to cover the entire Hilbert space.
We derived formulas relating the number of excited and defect bonds to the explicit spin arrangement in a given state, and we offered a detailed prescription for converting from the raw spin representation into the bond representation.
Most important, we argued that the population count for each kind of bond is a good quantum number for the model Hamiltonian considered in Sec. \ref{SEC:t-deformed-generalization}

Figure~\ref{FIG:phase-diagram-t-deformed} shows a summary of the zero-temperature quantum phase diagram, based on an extrapolation of various numerical measurements on finite-size systems to the thermodynamic limit. The diagonal measurements (those involving $\sigma^z$ only) are symmetric about the horizontal $(\gamma=0$) axis, whereas off-diagonal measurements (involving $\sigma^+$ and $\sigma^-$) are antisymmetric. The two-parameter extended model we have proposed is exactly solvable on the horizontal line ($\gamma=0$) and on the unit circle ($\eta^2 + \gamma^2 = 1$). For $r < 1$, the ferromagnetic ground state is doubly degenerate. For $r \ge 1$, the ground state is everywhere unique, except at the point $(r=1,\theta=180^\circ)$, for all system sizes satisfying $N= 0\ (\text{mod}\,4)$. Along the line $(r>1,\theta=180^\circ)$, the ground state is highly degenerate for sizes $N=2\ (\text{mod}\,4)$, since the desired doubly staggered pattern is prevented from forming; hence, these sizes are excluded.

The $t$-deformed model lies on the unit circle, and the line $\eta=0$ separates the gapped, ordered phases on the right from the gapless, disordered phases on the left. The latter belong to a region in which the excitation gap closes exponentially fast.
That unit circle also demarcates a boundary between regions that show ferromagnetic behavior inside and coexistence of antiferromagnetic and dimerized behavior elsewhere. The dimer order is strong in the upper and lower parts of the diagram, where quantum fluctuations are enhanced. On the other hand, the right part of the phase space favors z AFM behavior and the left supports z DSM behavior. The system is gapped everywhere except on ellipses inside the unit circle and on the boundary of the domes on the left part of the phase diagram. Most parts of the phase space favor the Fredkin-like Dyck-form ground state except the two domes and inside the unit circle. The two domes on the left and the Dyck-form $(\sztot = 0, N_\text{d} = 0)$ excitation on the top of the phase diagram both emerge only for $N \ge 10$; in the thermodynamic limit, they all touch the circle at $r \approx 1.7$, approaching it from opposite sides. Inside the two domes, the ground state favors $\sztot = 0$ and $N_\text{d}=1$, leaving Dyck-form to the higher energy state. Inside the unit circle, the ground state is ferromagnetic (higher $\lvert \sztot \rvert$) with no defects ($N_\text{d} = 0$). 

The tunable Hamiltonian we have proposed and studied in this paper puts the Fredkin model and its $t$-deformed generalization in the context of a larger space of models that have a well-define notion of conventional, excited, and defect bonds. This is interesting because the identification of the bond character relies on knowledge of the complete spin configuration; in other words it is a global rather than local property of the spin state and hence has a topological nature. Our work makes clear that, even though the $t$ deformation is frustration free, its quantum-disordered ground state is nonetheless a result of a special tuning of the competing interactions, one that carefully balances their ferromagnetic and antiferromagnetic tendencies.

%@@@@@@@@@@@@@@@@@@@@@@@@@@@@@@@@%@@@@@@@@@@@@@@@@@@@@@@@@@@@@@@@@
\appendix
\section{Generalization of the {\textit{\textbf{t}}}-deformed model}\label{Append:t-deformed-model}
We consider the colorless $S=1/2$ specialization of the frustration-free, $t$-deformed Fredkin spin chain described in Ref.~\onlinecite{Zhang-JMP-17}. 
The Hamiltonian
\begin{equation}
H(t) = H_F(t) + H_\text{boundary}
\end{equation}
is the sum of bulk and boundary terms,
\begin{equation}
H_F(t) = \sum_{j=2}^{N-1} \bigl(\ket{\phi^A_j} \bra{\phi^A_j} + \ket{\phi^B_j} \bra{\phi^B_j}\bigr)
\end{equation}
and 
\begin{equation}
H_\text{boundary} = \ket{\downarrow_1} \bra{\downarrow_1} +  \ket{\uparrow_N} \bra{\uparrow_N}.
\end{equation}
The operators in $H_F$ project onto the states
\begin{align}
\ket{\phi^A_j}  &= \frac{1}{\sqrt{1 + \vert t^A_j \vert ^2}}\left[  \ket{ \uparrow_{j-1}  \uparrow_j\downarrow_{j+1}} - t^A_j  \ket{ \uparrow_{j-1}  \downarrow_j \uparrow_{j+1}} \right]\\
\intertext{and}
\ket{\phi^B_j}  &= \frac{1}{\sqrt{1 + \vert t^B_j \vert ^2}}\left[  \ket{ \uparrow_{j-1}  \downarrow_j\downarrow_{j+1}} - t^B_j  \ket{ \downarrow_{j-1}  \uparrow_j \downarrow_{j+1}} \right].
\end{align}
The unspecified parameters satisfy $t^B_{j} = t^A_{j-1}$. If we treat them all on an equal footing, as in~Ref.~\cite{Deformed-spin-chain}, then the model depends on a single, site-independent parameter $t = t^A_j = t^B_j$.

The projector in Eq.~\eqref{EQ:fredkin_operator} can be represented as 
\begin{equation}\label{EQ:t-deformed}
P_{i,j} = \ket{S_{i,j}} \bra{S_{i,j}},
\end{equation}
where
\begin{equation}\label{EQ:singlet-pair}
\ket{S_{i,j}} = \frac{1}{\sqrt{2}}\left[\ket{ \uparrow_i \downarrow_j} - \ket{\downarrow_i \uparrow_j}\right]
\end{equation}
is the singlet formed by spin at sites $i$ and $j$. The multiparameter generalization~\cite{Zhang-JMP-17} of this state takes the form
\begin{equation}\label{EQ:t-deformed-singlet-pair}
\ket{S(t_{ij})_{i,j}} = \frac{1}{\sqrt{1 + t_{ij}^2}}\left[  \ket{ \uparrow_i \downarrow_j} - t_{ij}\ket{\downarrow_i \uparrow_j}\right],
\end{equation}
which is properly normalized and allows for admixing of the various spin-triplet components. The model preserves the frustration-free nature of the original Fredkin model in the sense that the ground state minimizes each term in the Hamiltonian individually.

For simplicity, we work on single-tuning-parameter $t$ deformation studied in Ref.~\onlinecite{Deformed-spin-chain},
\begin{equation}\label{EQ:t-deformed-singlet-pair}
\ket{S(t)_{i,j}} = \frac{1}{\sqrt{1 + t^2}}\left[  \ket{ \uparrow_i \downarrow_j} - t\ket{\downarrow_i \uparrow_j}\right],
\end{equation}
from which the usual Fredkin model is recovered at $t=1$.
The $t$-deformed ground consists of a sum of Dyck-form spin states whose weight is proportional to area under the corresponding height profile. Let us express the spin-singlet projector in terms of Pauli matrices, with $2\sigma^{\pm} = \sigma^x \pm i \sigma^y$ defining the raising and lowering operators. 
\begin{widetext}
\begin{equation}\label{EQ:t-deformed-singlet-projector}
\begin{split}
P(t)_{i,j} &= \ket{S(t)_{i,j}} \bra{S(t)_{i,j}} \\
     &=\frac{1}{1 + t^2}\left[ \left( \ket{ \uparrow_i \downarrow_j} 
        - t\ket{\downarrow_i \uparrow_j} \right)  \left( \bra{ \uparrow_i \downarrow_j} - t\bra{\downarrow_i \uparrow_j} \right)\right]\\
    &= \frac{1}{1 + t^2}\left[\ket{ \uparrow_i \downarrow_j} \bra{ \uparrow_i \downarrow_j} 
        + t^2 \ket{\downarrow_i \uparrow_j}\bra{\downarrow_i \uparrow_j}
    -t \left(  \ket{ \uparrow_i \downarrow_j}
          \bra{\downarrow_i \uparrow_j} + \ket{\downarrow_i \uparrow_j} \bra{ \uparrow_i \downarrow_j}\right)\right]  \\          
       &=\frac{1}{1 + t^2}\left[ \frac{1}{4}\left(\mathbb{1}+\sigma_i^z\right) \left(\mathbb{1}-\sigma_j^z\right) + \frac{t^2}{4} \left(\mathbb{1}-\sigma_i^z\right) \left(\mathbb{1}+\sigma_j^z\right) - t\left( \sigma_i^+\sigma_j^- + \sigma_i^-\sigma_j^+\right) \right] \\
          &=\frac{1}{1 + t^2}\left[\frac{1+t^2}{4} \left(\mathbb{1}-\sigma_i^z \sigma_j^z\right) + \frac{1-t^2}{4}\left(\sigma_i^z-\sigma_j^z\right)
          - t\left( \sigma_i^+\sigma_j^- + \sigma_i^-\sigma_j^+\right) \right] \\
& = \frac{1}{4}\left(\mathbb{1}-\sigma_i^z \sigma_j^z\right) +\frac{1-t^2}{4(1+t^2)}\left(\sigma_i^z-\sigma_j^z\right) -\frac{t}{1+t^2}\left( \sigma_i^+\sigma_j^- + \sigma_i^-\sigma_j^+\right) \\
  & \!\!\rightarrow \frac{1}{4}\left(\mathbb{1}-\sigma_i^z \sigma_j^z\right) +\frac{\eta}{4}\left(\sigma_i^z-\sigma_j^z\right) -\frac{\gamma}{2}\left( \sigma_i^+\sigma_j^- + \sigma_i^-\sigma_j^+\right) 
=: \tilde{P}_{i,j}(\eta,\gamma)
 \end{split}
\end{equation}
\end{widetext}
The final line is the two-parameter generalization, expressed as a function of $\eta$ and $\gamma$, that was introduced as Eq.~\eqref{EQ:generalized-projector}.
%@@@@@@@@@@@@@@@@@@@@@@@@@@@@@@@@@@@@@@@@@@@
 \bibliography{References}   % Use the BibTeX file ``References.bib''.
 
\end{document}